# Sub-nanosecond time resolution detector based on APD for Synchrotron Radiation ultrafast experiments

Li Zhen-jie(李贞杰), Li Qiu-ju(李秋菊), Liu Peng(刘鹏), Wang Shan-feng(王山峰),

Dong Wei-wei(董伟伟), Zhou Yang-fan*(周杨帆)

BSRF, Institute of High Energy Physics, CAS, Beijing 100049, China

zhouyf@ihep.ac.cn

**Abstract:** Synchrotron radiation light sources produce intense beam of X-ray with ultra-short pulse and nanosecond period. This offers the opportunities for the time resolution experiments[1]. Achieving higher counting rate and faster arriving time is difficult for common detectors. But avalanche photodiodes (APD) based on silicon which have been commercially available[1] with large active areas (e.g.10mm×10mm@ Perkin-Elmer Inc.) could satisfy the demands due to their good time resolution, low noise and large area[2-4]. We investigate the high counting rate and nanosecond time resolution detector with APD. The detector's fast amplifier was designed with the gain of about 60dB (1000). The amplifier included with three stages RF-preamplifier using MAR6+ chip[5] for the carefully controlling the circuit oscillation. Some measures have been taken for the preamplifiers good performance such as using resistance net between RF-preamplifier chip and the isolation of high voltage circuit from the preamplifier. The time resolution of the preamplifier together with APD sensor could reach below 1ns FWHM.

**Key words:** APD detector, Sub-nanosecond time resolution, Ultrafast experiment, Synchrotron radiation

**PACS:** 07.85.Qe, 29.20.dk, 29.40.Wk

## 1. Introduction

Avalanche photodiode (APD) is a kind of solid device used for direct X-ray detection for its good time resolution and high-dynamic range. The device, with single photon sensitivity and an intrinsically fast respond, is the detector of choice in variety of experiments, for example nuclear resonance scattering experiment, ultrafast experiment of synchrotron radiation. This paper focus on the time resolution based on large area APD of C30703 (10mm×10mm@ Perkin-Elmer Inc.) with gain of 60 dB RF-preamplifier. Firstly, we review the basic properties and principles of APD devices. For the X-ray detection, we concern the efficiency of photon detection. Then the electronic with 60dB gain is described, and we give the notice of the design of high gain amplifier without oscillation. Finally we discuss the properties of the detector system, test the time resolution with different energies and study the responses to the different X-ray photon energies.

## 2. Overview of APD

In the last few years, semiconductor detectors have been used for X-ray and $\gamma$ ray detection for their good performance (working at room temperature, high resolution, high counting rate and fast response time). Among the semiconductor detectors, APD is a kind of special one with internal signal amplification. Such an 'internal gain', caused by the electron cascades in the high electric field, usually about $10^5$ V/cm, improves the performance of signal-to-noise[6]. For the reason of high electric field, the signal in the APD can be rapidly collected by the electrode (from tens of picoseconds to few nanoseconds with different designed APD)[7]. The time resolution of APD has been pushed as low as 20ps when cooled and operated in Geiger mode[8-9].

There are three types of APDs available for photon detection: beveled edge, reverse and reach-through diode. These three types APDs have different properties depending on their different structure. A schematic diagram is shown in Fig.1. The beveled-edge APD structure is shown in Fig.1 (a). They are fabricated by performing a deep p-type diffuse into a highly resistive n-type substrate, yielding a uniform junction with a low built-in electric field[10]. The beveled-edge APDs have the advantages of high gain, low dark current, and low excess noise factor for the broad multiplication region. However the disadvantages of beveled-edge APDs are high operating voltage (typically greater than 1000V) and limited rise time. For the broad multiplication region, the beveled-edge APDs are inconvenient for X-ray spectroscopy. When the X-ray photons deposit in the silicon buck, the collected charges depends on the position where the photons are absorbed. So, the signal pulse height could not reflect the energy of the incoming photons. The gain variation results an undesired peak background in the real spectrum.

Fig.1 (b) shows the structure of reverse type APDs with low noise for the reason that the multiplication region is just behind the detector surface. Only the surface dark current is amplified and the buck dark current is not. This kind of APD is design to couple to scintillators. The narrow multiplying region is in front of the incoming window, typically about 5um from the surface of the device[11]. Most of the dark current undergoes only the holes multiplication, so it reduces the noise and has the advantage of low energy detection.

The reach-through APDs shown in Fig.1(c) contain a narrow multiplying region similar to reverse type APDs consisting of p-type and n-type layers. The depletion region is separated in two regions: a thick drift region and a multiplying region. The multiplying region is usually designed to few micrometers and the drift region occupies more than 90% of the APD total volume. The X-ray photons are absorbed in this region, interacting with the semiconductor and creating electron-holes. The electrons move toward the multiplying region. All the photons absorbed in the drift region could create the similarly signal pulse height[12].

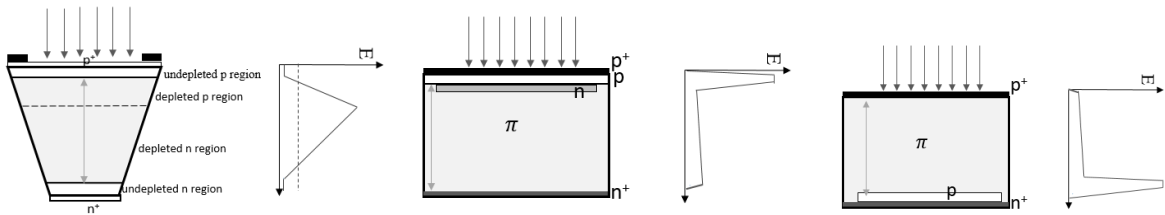

Fig.1 (a) The beveled-edge APD (b) reverse type APD (c) reach-through APD. The electric field profiles in each APD are shown in right panels.

In this paper, Perkin-Elmer C30703 reach-through type APDs are used in our detector system for the time and energy detection. The C30703 APDs offer the available combination of high speed, low noise and capacitance and extended IR response. The thickness of the silicon wafer is 110um, sufficient enough to detect soft X-ray (6keV-15keV) and the efficiency at different energies was shown in fig.2. The efficiency could reach 20% at the X-ray energy of 15keV. The main properties of C30703 APDs are summarized in Table 1.

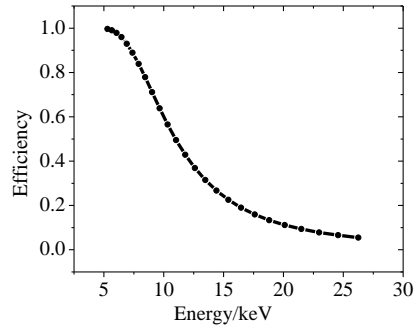

Fig.2 The efficiency of silicon detector as the X-ray energy with the thickness of *110 um*

Table 1

Properties of reach-through type APD C30703FH

| Parameters of Perkin-Elmer C30703FH reach-through type APDs | |
|---|---|
| Photo Sensitive Diameter | 10mmX10mm |
| dark current (M=50) | 10nA |
| Response time | 5ns |
| Vop range | 275V-425V |
| Capacitance | 120pF |
| Ionization Ratio (k) | 0.02 |
| Dark noise ($i_n$) | 0.7pA/$\sqrt{Hz}$ |

## 3. Electronics -Amplifier for APDs and Signal from the amplifier

The signal from the APDs is not big enough for the next measurement, so the amplifier should be used for the APDs. The electronic circuit used in this paper is shown in Fig.3. The circuit contains two parts: the high voltage control and signal amplication. APDs are kept at positive (or negative) high voltage. The high voltage is applied in the APDs through a high value resistance. This high value resitance (usually from few hundred kohm to few Mohm) could protect the APDs from flowing through big current. Once the photons are absorbed in the silicon, the signal from the APDs is feed to the high gain amplifier circuit. Here we use MAR6+ amplifier chip, cascading 3 of them sequentially for a total excepted gain of about 60dB. The amplifier voltage gain shold be above 500 for a modest amplititude signal. But this high voltage gain circuit is difficult to get and some care is needed to insure the amplifier operated without oscillation. In our experiment, we add the resistance net between two amplifier chips to improve the oscillation. However, this resistance net would reduce the gain a little[13].

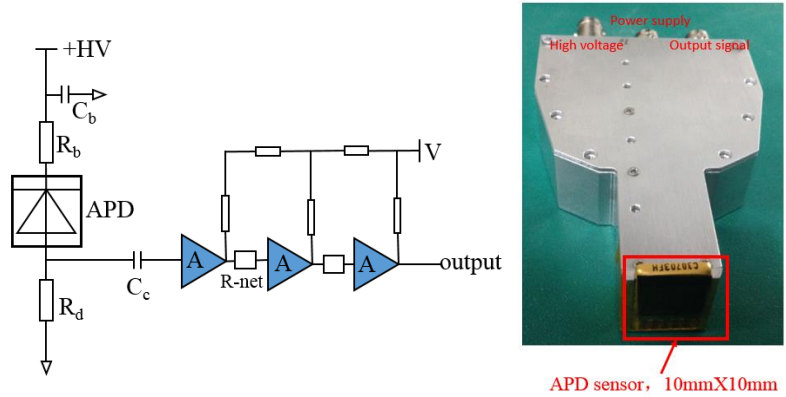

Fig. 3 Left: Schematic of Amplifier electronics with 3 stages RF-preamplifier used in our system. Rb (470kOhm) limits the bias current and Rd (27kOhm) is closed to the bias circuit. The pulse of the fast signal from APD goes through Cc to the first stage amplifier and to the ground. This is the low impedance (50Ohm) path saw by the signal. The circuit is hand mounted on the circuit board with a copper ground plane. Right: the APD detector

The measurements of the directct single X-rays take palce in BSRF-1W2B. Fig.4 shows a typical signal shape of a single X-ray photons and the noise line of the pre-amplifier. The full pulse width of the signal is about less than 30ns and the signal leading edge is less than 5ns. So the APD detector could work in high count rate (above 20MHz, $2\times10^7$ count rate). From Fig.4, the noise baseline of the APD pre-amplifier system is about $\pm50$mV, and the amplitude of X-ray photon signal is above 100mV.

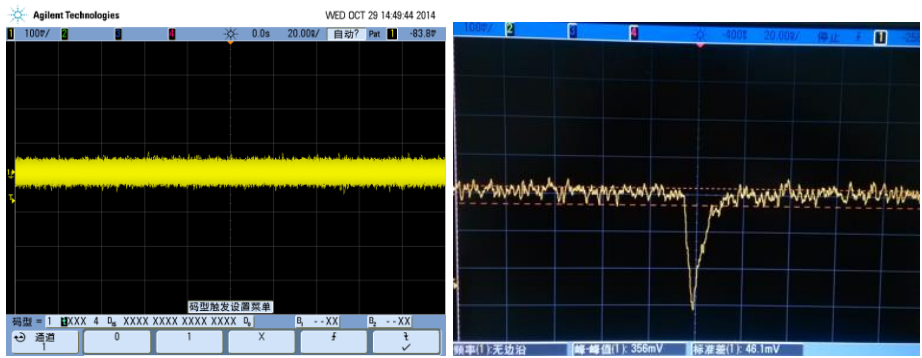

Fig.4 (a) The noise lever of APD detector output both with APD sensor and the preamplifier, and the maximum peak-to-peak noise value is about 100 mV. (b) The signal shape of single photon with the pulse width of 20ns.

### 4. Time resolution and Application in Nuclear Resonance Scattering Experiment

The time resolution of the APD was tested using single bunch beam mode as discribed in Fig.5 and Fig.6. The single X-ray beam was created by the single electron bunch which is seperated from other bunches by several hundred nanoseconds. The timing signal is created by the accelerator RF signal with picosecond precision. The timing signal is precise enough to measure the time resolution of nanosecond as the start signal.

The X-ray photons were absorbed by the APD detector with fast signal output from the preamplifier. The fast signal was then send to the oscilloscope to measure the different value between the timing leading edge signal and the photon signal. The threashold (above noise value) should be set for the photons signal to avoid the influence of the noise. The experiment setup is sketched in Fig.6.The single photon bunch created by single electron bunch enters the APD, and the time between the timing and the fast signal was measured via the oscilloscope. We measured the time resolution at different energies and the results were summarized in Fig.7 and Table 2. The FWHMs of time resolution at different energies were all within 1 nanosecond, about 850 picoseconds and the sigmas of the resolution were about 360 picoseconds.

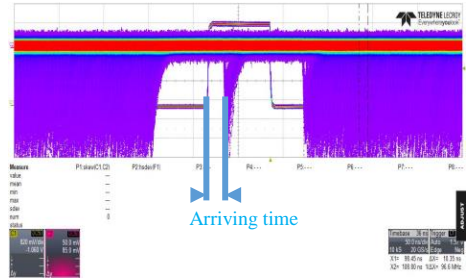

Fig.5 The single bunch of electron was chosen by the timing signal and the time between the timing leading edge signal and the photon leading edge signal was measured.

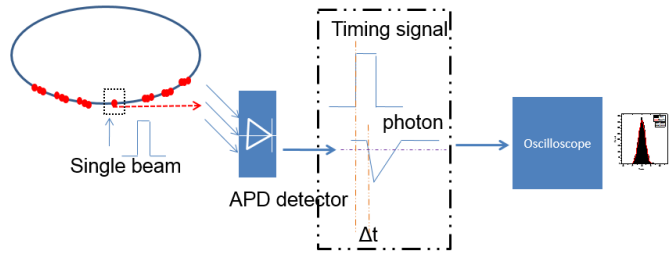

Fig.6 Experiment setup for the time resolution measured. The X-ray photons created by the single electron beam entered in the APD, the timing signal was precise enough to be set as the reference signal and the Δt was measured by the oscilloscope

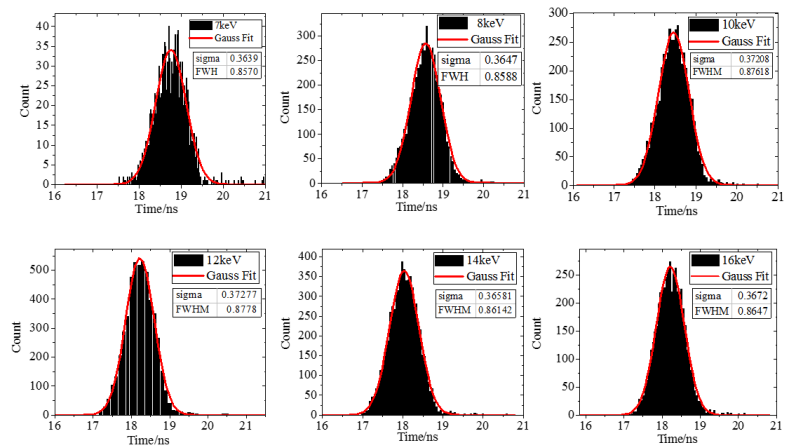

Fig.7 Time resolution at diffetent energies from 7keV to 16keV, and the Gauss fittings were add to every statistical data.

Table 2

Time resolution at different energies

| Photon Energy | FWHM(ns) | Sigma(ns) |
|---|---|---|
| 7.0 keV | 0.857 | 0.364 |
| 8.0 keV | 0.859 | 0.364 |
| 10.0 keV | 0.876 | 0.372 |
| 12.0 keV | 0.878 | 0.372 |
| 14.4 keV | 0.861 | 0.365 |
| 16.0 keV | 0.865 | 0.367 |

Timing resolution of the APD, assuming the noise of the system is zero, is limited by the location of the absorption of the X photons and the transit time of the carrier in the APD drift region. The Fig.8 shows at room temperature the maximum delay (in other words the photon is absorbed on the opposite side as the gain region (junction)). This delay will be uniformaly distributed from 0 up to this value show in the graph. For instance, assuming 120um APD, at 350V, the maximum delay is 1.80ns. Since the delay will be uniformly distributed, the timing resolution (standard deviation) will be sqrt(1.80ns/12) = 380ps or so. If the X photon was absorbed in the body, the delay will be short than the maximum delay time and the time resolution will be get better. In our experiment, the X-ray photon will enter in the APD body and distribute from the surface to the junction. So, the delay time of every single photon is shorter than the maximum delay time, and the time resolution will be better than 380ps. This result accords with our measurements.

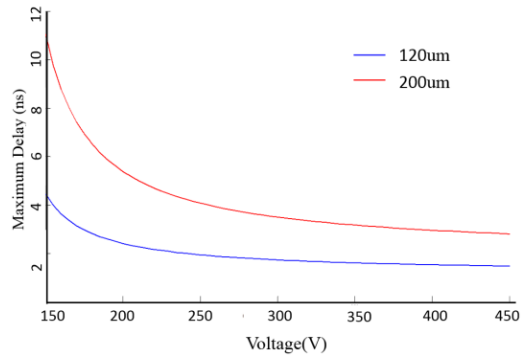

Fig.8 The maximum delay time of APD sensor at different bias voltage with different sensor thickness. The delay time is corresponding to the intrinsic time resolution of the APD sensor, the timing resolution (standard deviation) will be sqrt(delay time/12).

We tested the APD detectors in APS-3-ID beamline and the $^{57}$Fe spectra of nuclear resonance scattering was measured in Fig.9. In this figure, the delay time signals from 10ns to 130ns were recored clearly and the subnanosecond time resolution was easily get. The APD_1 and APD_2 showed the same trend and they worked robust during the experiment.

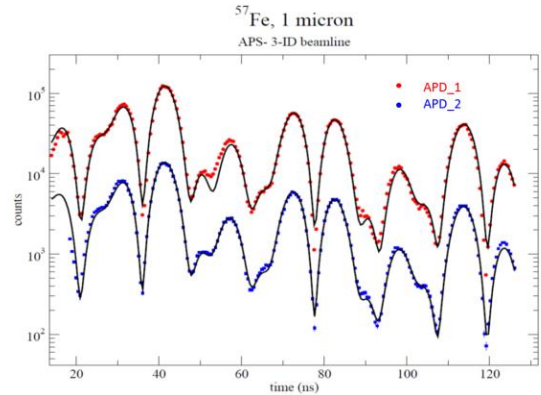

Fig.9 The $^{57}$Fe spectra of nuclear resonance scattering measured in APS-3-ID beamline

### 5. Conclution

We have studied the performance of large area APDs developed by Perkin-Elmer Inc and presented a clear picture of time respoonse. This paper showed the reach-through APDs of C30703FH with the thickness of 110um could get excellent time resolution. The FWHM of the time resolution was about 850 picoseconds and the sigma was about 360 picoseconds. The results showed the time resolutions were similar at different energies. APDs have faster response and larger dynamic range used for direct detection of X-rays. The time resolution could be get excellent with thicker and smaller size sensor for the reduced capacitance. The APD detectors have been used for ultrafast time resolution experiment in BSRF and nuclear resonance scattering experiment in APS (Advanced Photon Source in America).